\documentclass[aps,prd,preprint,superscriptaddress,tightenlines,nofootinbib,showpacs]{revtex4}

\usepackage{graphicx}%
\usepackage{dcolumn}%
\usepackage{bm}
\newcommand{ \mbc }{ $M_{\rm bc}$ }
\newcommand{\kone}{K_{\rm 1}({\rm 1270})}
\newcommand{\konem}{K^-_{\rm 1}({\rm 1270})}
\newcommand{\kones}{K_{\rm 1}({\rm 1400})}
\newcommand{\gevcsqsq}{(\rm{GeV}/$c$^2)^2}
\begin{document}

\preprint{CLNS 07/1994}      
\preprint{CLEO 07-4}     
    
\title{Evidence for the Decay $D^0 \rightarrow K^- \pi^+ \pi^-e^+ \nu _e$}
\author{M.~Artuso}
\author{S.~Blusk}
\author{J.~Butt}
\author{J.~Li}
\author{N.~Menaa}
\author{R.~Mountain}
\author{S.~Nisar}
\author{K.~Randrianarivony}
\author{R.~Sia}
\author{T.~Skwarnicki}
\author{S.~Stone}
\author{J.~C.~Wang}
\author{K.~Zhang}
\affiliation{Syracuse University, Syracuse, New York 13244}
% ...

\collaboration{CLEO Collaboration} 
\noaffiliation

\date{May 29, 2007}

\begin{abstract}
Using a 281 pb$^{-1}$ data sample collected at the $\psi(3770)$ with
the CLEO-c detector, we present the first absolute branching
fraction measurement of the decay $D^0 \rightarrow K^-\pi^+\pi^-
e^+\nu _e$ at a statistical significance of about 4.0 standard
deviations. We find 10 candidates consistent with the decay $D^0
\rightarrow K^-\pi^+\pi^- e^+\nu _e$. The probability that a
background fluctuation accounts for this signal is less than $4.1
\times 10^{-5}$. We find ${\cal B}(D^0 \rightarrow K^-\pi^+\pi^-
e^+\nu _e)= [2.8 ^{+1.4}_{-1.1}{\rm (stat)} \pm 0.3{\rm (syst)}]
\times 10 ^{-4}$. This channel is consistent with being
predominantly produced through $D^0 \rightarrow \konem\ e^+\nu _e$.
By restricting the invariant mass of the hadronic system to be
consistent with $\kone$, we obtain the product of branching
fractions ${\cal B}(D^0 \rightarrow \konem\ e^+\nu _e)\cdot{\cal
B}(\konem\to K^-\pi^+\pi^-)=[2.5^{+1.3}_{-1.0}\ {\rm(stat)} \pm 0.2
\ {\rm (syst)}]\times 10^{-4}$. Using ${\cal B}(\konem\to
K^-\pi^+\pi^-) = (33\pm 3) \%$, we obtain ${\cal B}(D^0 \rightarrow
\konem\ e^+\nu _e)=[7.6^{+4.1}_{-3.0}\ {\rm (stat)} \pm 0.6\ {\rm
(syst)}\pm 0.7 ]\times 10^{-4}$. The last error accounts for the
uncertainties in the measured $\konem\to K^-\pi^+\pi^-$ branching
fraction.


\end{abstract}


\maketitle


The understanding of the hadronic mass spectrum in semileptonic
decays of charm mesons sheds light on non-perturbative strong
interaction dynamics in weak decays. In particular, an interesting
question is whether the charm quark can be considered ``heavy,'' and
thus theoretical predictions based upon heavy quark effective theory
(HQET) can be applied to describe some features of its decays. A
priori this seems to be an unlikely scenario as, even in the
Cabibbo-favored transition $c \to s e^+ \nu_e$, the daughter quark
is too light for HQET to apply. Nonetheless, this effective theory
seems to describe these decays relatively well~\cite{isgw2}.

The decays induced by the quark level process  $c \rightarrow s e^+
\nu_e$ are dominated by the two final states $D\to K e^+ \nu_e$ and
$D\to K^\ast e^+ \nu_e$. CLEO-c has measured exclusive $D$
semileptonic branching fractions for all modes observed to date: $K
e^+ \nu_e$, $K^* e^+ \nu_e$, $\pi e^+ \nu_e$, $\rho e^+ \nu_e$, and
$D^+ \rightarrow \omega e^+ \nu_e$~\cite{DExcl_57invpb}, as well as
inclusive $D \rightarrow X e^+ \nu_e$ branching
fractions~\cite{cleo-c-incl}. The sum of the exclusive branching
fractions and the inclusive branching fractions for $D$ meson
semileptonic decays are consistent: $\sum \mathcal{B}(D^0_{\rm
excl}) = [6.1 \pm 0.2{\rm (stat)} \pm 0.2{\rm (syst)}]$\% and $\sum
\mathcal{B}(D^+_{\rm excl}) = [15.1 \pm 0.5{\rm (stat)} \pm 0.5{\rm
(syst)}]$\%  while $\mathcal{B}(D^0 \rightarrow X e^+ \nu_e) = [6.46
\pm 0.17{\rm (stat)} \pm 0.13{\rm (syst)}]$\% and $\mathcal{B}(D^+
\rightarrow X e^+ \nu_e) = [16.13 \pm 0.20{\rm (stat)} \pm 0.33{\rm
(syst)}]$\%. Nonetheless, there is some room left for higher
multiplicity modes.

The quark model developed by Isgur, Scora, Grinstein, and Wise
\cite{isgw}, later updated to include constraints from heavy quark
symmetry, hyperfine distortions of wave functions, and form factors
with more realistic high recoil behavior \cite{isgw2}, is the only
one to provide quantitative predictions for the partial width of
decays such as $D\rightarrow \kone\ e^+\nu_e$. In general, we expect
the decay mediated by the quark level process $c\to se^+\nu_e$ to be
dominated by the ground state pseudoscalar and vector daughter
mesons. The low available phase space makes it less likely to
produce heavier mesons, such as $P$-wave or first radial excitations
of the $s\bar{u}$ and $s\bar{d}$ quark states. The lightest excited
state is the $\kone$. This model predicts that the partial width
$\Gamma(D\to \kone e^+\nu_e)$ is 2\% of the total $\Gamma(c\to s
e^+\nu_e)$, and that decays to other excited resonances are
suppressed by at least a factor of 10 more.

Little is known about $D^0\rightarrow \kone e^+\nu _e$ to date. The
fixed target experiment E653 \cite{esixfivethree} reported a 90\%
confidence upper limit of ${\cal B} (D^0 \to K^-\pi^+\pi^-\mu^+\nu
_{\mu})< 0.037 \times {\cal B} (D^0 \to K^-\mu^+\nu _{\mu})$. This
Letter is the first report on a signal for the decay $D^0 \to
K^-\pi^+\pi^-e^+\nu _e$.

We use a 281 pb$^{-1}$ data sample collected at the $\psi({\rm
3770})$ with the CLEO-c detector \cite{cleoiii, cleoc}. The three major
subsystems of this detector are the charged particle tracking
chambers, the CsI electromagnetic calorimeter, and a Ring Imaging
Cherenkov (RICH) charged particle identification system. All these
components are critical to an efficient and highly selective
electron and positron identification algorithm. The CsI calorimeter
measures the electron and photon energies with an r.m.s. resolution of 2.2\%
at $E=1$~GeV and 5\% at $E$=100~MeV. One of the key  variables for
$e$ identification, $E/p$, uses $E$, the energy measured in the
calorimeter and $p$, the momentum measured in the charged particle
tracking system. The tracking system is composed of a 6-layer
inner drift chamber and a 47-layer main drift chamber. The main drift
chamber also provides specific ionization ($dE/dx$) measurements
for charged particle identification. In addition, charged
particles are identified using the
RICH detector \cite{rich}.  Combining information from these
detector subsystems, we achieve efficient and selective charged
particle identification over the entire momentum region relevant for
the decays studied.

\begin{figure}[htbp]
\includegraphics*[width=3.5in]{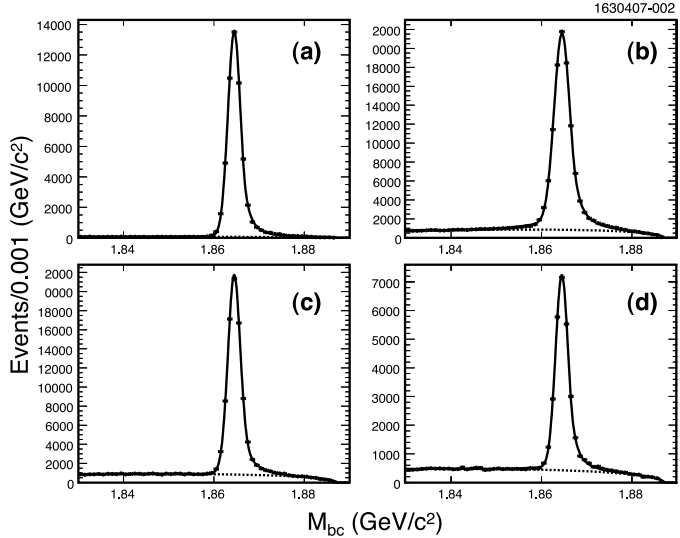}
\caption{\small $M_{\rm bc}$ spectra for (a) $\bar{D}^0\to
K^{+}\pi^{-}$, (b)$ \bar{D}^0\to K^{+}\pi^{-}\pi^{0}$, (c)
$\bar{D}^0\to K^{+}\pi^{-}\pi^{+}\pi^{-}$, (d)$\bar{D}^0\to
K_{S}^{0}\pi^{+}\pi^{-}$ candidate tags.} \label{fig:tags}
\end{figure}

We use a tagging technique similar to the one pioneered by the Mark
III collaboration \cite{markIII}. Details on the tagging selection
procedure are given in Ref.~\cite{dhad}. We select events containing
a fully reconstructed $\bar{D}^0 \rightarrow K^+\pi^-$, $\bar{D}^0
\rightarrow K^+\pi^-\pi^0$, $\bar{D}^0 \rightarrow
K^+\pi^-\pi^+\pi^-$, or $\bar{D}^0 \rightarrow K_{S}^{0}\pi^+\pi^-$
decay, which we call a tag. (Mention of a specific mode implies the
use of the charge conjugate mode as well throughout this Letter).
Two kinematic variables, namely energy difference, $\Delta E \equiv
E_{\text{tag}}- E_{\rm beam}$, and beam-constrained mass, $M_{\rm
bc}\equiv \sqrt{E_{\rm beam}^2/c^4-
 |\vec{p}_{\text{tag}}|^2/c^2}$, are used to select tag candidates, where
$E_{\rm beam}$ represents the beam energy and $(E_{\text{tag}},\vec{p}_{\text{tag}})$
represent the 4-vectors of the $\bar{D}^{0}$ tag candidate. We first require
$|\Delta E|$ to be less than 0.020 to 0.030 GeV, depending upon the
mode considered.   Figure~\ref{fig:tags} shows the $M_{\rm bc}$
spectra for events that satisfy the $|\Delta E|$ requirement for the
four tagging modes considered. In order to determine the total
number of tags, we fit the \mbc\ distribution with a signal shape
composed of a Crystal Ball function \cite{cball} and a Gaussian, and
an ARGUS function \cite{argus-fit} parameterizing the background in the fit.
The signal window is chosen as 1.858~GeV/$c^2 \le
M_{\rm bc} \le 1.874$~GeV/$c^2$. In order to extract the tag yield,
we integrate the signal shape within this \mbc\ interval.
Alternatively, we count tag candidates in the $M_{\rm bc}$ signal
window and subtract the combinatorial background obtained by
integrating the background function from the fit. The total number
of tags obtained with the former method is $[257.4 \pm 0.7 {\rm
(stat)}] \times 10^{3}$; the second method gives $[257.7\pm 0.6 {\rm
(stat)}] \times 10^{3}$. The agreement is excellent and we use the
latter number as the total number of tags in our sample. The
difference between the two tag yields is included in a systematic
uncertainty.

In each event where a tag is found, we search for a set of tracks
recoiling against the tag that are consistent with a semileptonic
decay. We select tracks that are well-measured and have a helical
trajectory approaching the event origin within a distance of 5~cm
(5~mm) along the beam axis (in the plane perpendicular to the beam
axis). Each track must include at least 50\% of the main drift chamber
wire hits expected for its momentum and have momentum greater than
50 MeV/$c$. We search for a positron among well reconstructed tracks
having a momentum of at least 200 MeV/$c$, as the electron identification
 becomes increasingly difficult at low momenta. We also
require $|\cos{\theta}| < 0.90$, where $\theta$ is the angle between
the positron direction and the beam axis. The positron selection
criteria are discussed in Ref.~\cite{DExcl_57invpb}. They have an
average efficiency of 95\% in the momentum region $[0.3 - 1.0]$
GeV/$c$, and 71\% in the region $[0.2 - 0.3]$ GeV/$c$. In addition,
we search for a good track consistent with a $K^-$ and two
oppositely charged tracks consistent with pions. Hadron track
identification criteria rely on $dE/dx$ information from the drift
chamber for tracks with $p<0.7$ GeV/$c$. For tracks with $p \ge
0.7$~GeV/$c$, in addition to $dE/dx$ measurements, information from
the RICH detector \cite{rich} is used to improve the $K$-$\pi$
discrimination. In the momentum range relevant for this analysis the
$K$-$\pi$ misidentification probability is negligible. The $e$-$\pi$
misidentification probability, determined experimentally with
radiative Bhabhas, has an average value of 17\% for electron momenta
below 0.2 GeV/$c$, and is about 1\% for higher momenta.

As the decay mode that we are investigating is rare, efficient
background suppression is critical to achieve adequate sensitivity.
Accordingly, we require that only four charged tracks be present in
the event in addition to those used in the tag reconstruction. The
dominant source of background in this analysis arises from events in
which the detected positron comes from a $\gamma$ conversion
($\gamma \to e^+e^-$), or a $\pi^0$ Dalitz decay ($\pi^0 \to
e^+e^-\gamma$). This background is equally likely to produce $(e^+
K^-)$ combinations, which we call right-sign events~(RS), and $(e^-
K^-)$ combinations, which we call wrong-sign events~(WS). Typically,
an $e^+e^-$ pair arising from a conversion $\gamma$ or a $\pi^0$
Dalitz decay has a strong angular correlation with almost collinear
angular orientation of the two particles. For signal events, the
opening angle between the $e^+ \pi^-$ pair tends to be large.  We
therefore include a requirement that the opening angle be greater
than $20 ^\circ$.  This requirement eliminates most of the
background from conversion $\gamma$'s or $\pi^0$ Dalitz decays,
while reducing the signal efficiency by only 1.7\%.

\begin{figure}[htbp]
\includegraphics*[width=3.5in]{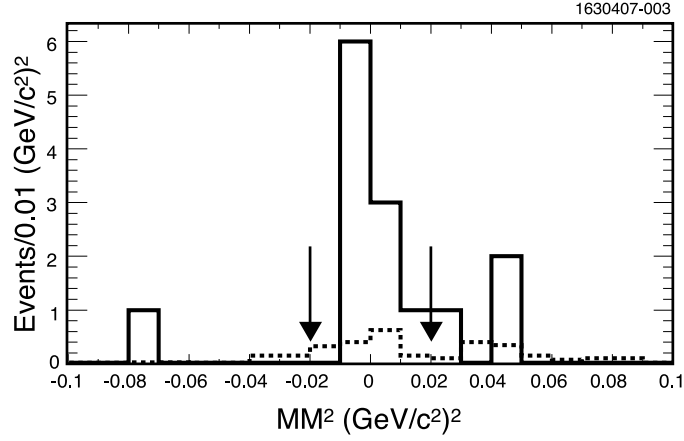}
\caption{\small Missing mass squared ($MM^2$) distribution for the
RS sample $D^0\to K^-\pi^+\pi^- e^+\nu_e$. The dashed histogram
represents the estimated background. Events with $MM^2$ within the
two arrows are considered signal candidates.}\label{fig:MM2}
\end{figure}

In this semileptonic sample, signal candidate events are selected
using the missing mass squared $MM^2$ defined as
\begin{equation}
MM^2 = (E_{\text{beam}}-\sum_{i=1}^{4}
E_i)^2/c^4-(-\vec{p}_{\text{tag}}-\sum_{i=1}^{4}\vec{p}_i)^2/c^2,
\end{equation}
where  $\vec{p}_{\text{tag}}$ is the momentum of the fully
reconstructed tag, and $(E_i,\vec{p}_i)$ represent the energy and
momentum of the four tracks in the $D^0$ candidate. For signal
events the $MM^2$ distribution is centered at zero, as it represents
the invariant mass squared of the missing $\nu_e$. According to
Monte Carlo simulation of our signal semileptonic channel, the $MM^2
$ distribution has a resolution consistent among the tag modes with
a standard deviation~($\sigma$) of $0.00594 \pm 0.00010\ \gevcsqsq$.
Figure~\ref{fig:MM2} shows the measured $MM^2$ distribution for RS
events in the data as well as the estimated background, derived from
GEANT-based Monte Carlo simulation \cite{geant3} in combination with
particle misidentification probabilities derived from data. In
addition we estimate the background directly from the WS events in
data. We define a signal window as $|MM^2| \le 0.02\ \gevcsqsq$.
There are 10 events in the signal window of $MM^2$ as shown in
Figure~\ref{fig:MM2}.

\begin{figure}[htbp]
\includegraphics*[width=3.5in]{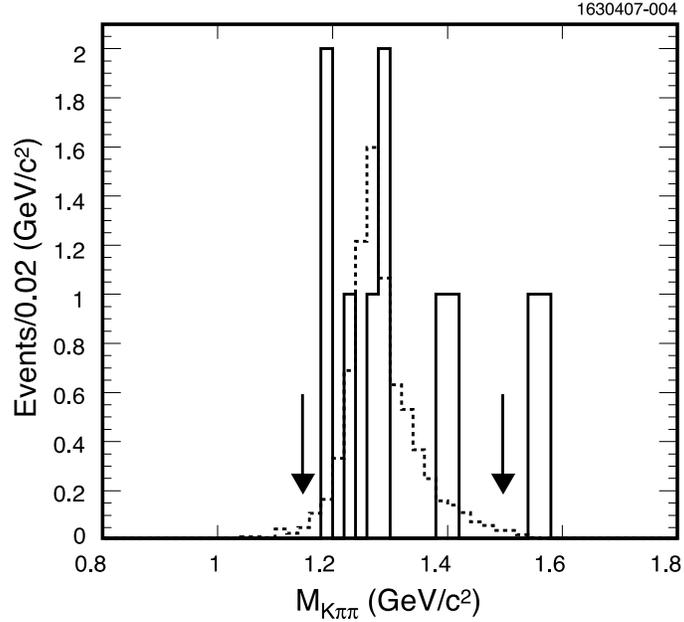}
\caption{\small Invariant mass of the hadronic system in the data
for $D^0\to K^-\pi^+\pi^- e^+\nu _e$. The dashed histogram
represents the predicted distribution obtained using a Monte Carlo
simulation according to the ISGW2 model, assuming all the
$K^-\pi^+\pi^-$ are $\kone$ decay products. The region within the
two arrows defines the invariant mass range used to select the
$\konem$ resonance.} \label{fig:mhad}
\end{figure}

Another interesting observable is the invariant mass of the
$K^-\pi^+\pi^-$ hadron system. Figure~\ref{fig:mhad} shows the
invariant mass of the $K^-\pi^+\pi^-$ system for RS candidate
events, compared with the expectation from the ISGW2 model
\cite{isgw2}, which provides the best representation of our data,
where the hadronic system forms the $\kone$ resonance. The
measured distribution is in reasonable agreement with this model.

We have performed several studies to determine possible background
sources. A Monte Carlo sample incorporating all the information
available on $D$ meson decays  and 40 times bigger than our
collected data demonstrates that the dominant background comes from
conversion $\gamma$'s or $\pi^0$ Dalitz decays.  As the $e$ to $\pi$
misidentification probability may not be modeled accurately by our
Monte Carlo simulation, the background from Dalitz decays is
evaluated by folding the $e$ spectra from simulated $D^0\rightarrow
K^-\pi^+\pi^0$ decays with the $e$ to $\pi$ misidentification
probability derived from a radiative Bhabha data sample. This study
predicts that 1.56 $\pm$ 0.22 background events are due to this
source if no requirement on the $K^-\pi^+\pi^-$ invariant mass is
applied.  A study of the WS data gives one background event, in
agreement with the previous estimate. In addition, there are small
background components from the decays $D^0\to K^-\pi^+\pi^+\pi^-$
($0.2\pm 0.1)$ and $D^0\to K^-\pi^+\pi^+\pi^-\pi^0$ ($0.1\pm 0.1)$,
both estimated with Monte Carlo samples. We have also studied
non-$D\bar{D}$ contributions at this center-of-mass energy, such as
those from the continuum ($e^+e^- \to q \bar{q}$, where $q$ is a
$u$, $d$, or $s$ quark), radiative return production of $\psi(2S)$,
and $e^+e^-\rightarrow \tau ^+\tau ^-$ processes, and we do not find
any background from these sources. Summing up all background
contributions, we find that $1.86\pm 0.25\ {\rm (stat)}$ events are
consistent with background. We have also studied this sample with a
requirement on the invariant mass of the $K\pi\pi$ system
 optimized for the decay $D^0\to \konem e^+ \nu _e$, using the variable
 $S/\sqrt{S+B}$, where $S$ is the number of signal events predicted
 from Monte Carlo simulations and $B$ is the number of estimated background events. For the optimal
 invariant mass interval,
$[1150 - 1500]$ MeV/$c^{2}$, we find 8 candidate events and an
estimated background of $1.0^{+0.4}_{-0.3}{\rm (stat)}$ events, with
no events in the WS sample.

The reconstruction efficiency depends on the invariant mass  of the
$K^-\pi^+\pi^-$ system ($M_{\rm had}$).  A larger fraction of the
electron spectrum is below the momentum cut of $0.2$ GeV/$c$ for
higher $M_{\rm had}$, and the spin and parity of the final hadronic
state influence the electron spectrum shape as well. For example,
the ISGW2 model studies all the $P$-wave $s\bar{u}$ and $d\bar{u}$
hadronic final states, as well as the corresponding radial
excitations. Among the $P$-wave states, the $^{3/2}{P}_1$ are
identified with the $\kone$, and $^{1/2}{ P}_1$ are identified with
the $\kones$. The latter has a much softer electron spectrum, and
therefore our efficiency for detecting it is smaller. We have
studied the signal reconstruction efficiency with the ISGW2 model,
including different mixing percentages of the $^{3/2}{P}_1$ and
$^{1/2}{P}_1$ final states, as well as a phase space model for the
distribution of the $M_{\rm had}$. With the Monte Carlo simulation
based on the ISGW2 model,
 we obtain $\epsilon = (10.78\pm 0.23)\%$ for the full
$M_{\rm had}$ range and $\epsilon = (10.53 \pm 0.22)\%$ with the
$\kone$ mass requirement ($[1150 - 1500]$ MeV/$c^{2}$).

The absolute branching fraction for $D^0\rightarrow
K^-\pi^+\pi^-e^+\nu_e$ is obtained using $\mathcal{B} \equiv (N_{\rm
s} - N_{\rm b}) / (\epsilon_{\rm eff} N_{\rm tag})$,
where $N_{\rm s}$ is the number of signal events, $N_{\rm b}$ is the
number of background events, $N_{\rm{tag}}$ is the number of tags,
and $\epsilon_{\rm eff}$ is the effective efficiency for detecting
the semileptonic decay in an event with an identified tag. This effective efficiency
includes a correction term $C\equiv\epsilon_{\rm tag}^{\rm
sl}/\epsilon_{\rm tag}$ accounting for the small difference in tag
reconstruction efficiency in events containing the semileptonic
signal and in generic $D\bar{D}$ events. The average value of $C$ is
1.036. We obtain ${\cal B}(D^0\rightarrow K^-\pi^+\pi^- e^+\nu
_e)=(2.8 ^{+1.4}_{-1.1}\pm 0.3)\times 10^{-4}$, without applying any
invariant mass requirement. If we apply the $\kone$ invariant mass
requirement, we obtain ${\cal B}(D^0 \rightarrow \konem e^{+} \nu
_{e}) \cdot {\cal B}(\konem\rightarrow K^-\pi^+\pi^-) = (2.5^{+1.3}_{-1.0} \pm
0.2) \times 10^{-4}$. The smaller systematic uncertainty is derived by
the fact that the model dependence can simply be estimated by
varying the form factors in the ISGW2 model. In this case we do not need
to model
a broader invariant mass distribution for the $K^-\pi^+\pi^-$
system. Note that the probability for 1.86 background events to
fluctuate to 10 or more events, taking into account a
0.25 event Gaussian uncertainty, is $4.1 \times 10^{-5}$, corresponding  to a
 significance
of about 4.0 $\sigma$. The result with the $K^-\pi^+\pi^-$ mass
requirement has similar statistical significance.

\begin{table}[htpb]
\begin{center}
\vskip 0.5 cm
\begin{tabular}{lcc}
\hline \hline ~~& Systematic errors (\%)\\ \hline Number of tags &
0.5 & 0.5 \\
Tracking & 1.3 & 1.3\\
PID Efficiency (hadrons) & 1.9 &1.9 \\
PID Efficiency (electrons) & 1.0& 1.0\\
Opening angle cut & 1.5 & 1.5\\
$M_{K\pi\pi}$ cut  & -- & 1.7  \\

Model dependence & 10.0 & 4.0 \\
Background & 5.3  & 5.3\\
\hline Total & 11.9& 7.5 \\\hline\hline
\end{tabular}
\caption{\label{tab:sys}  Systematic errors on $D^0\to
K^-\pi^+\pi^-e^+\nu _e$ branching fraction. The first column applies to
the analysis without $\kone$ mass cut, the second to the analysis
with the $\kone$ mass cut as described in the text.}
\end{center}
\end{table}

The systematic uncertainties for the branching fractions are listed
in Table~\ref{tab:sys} and are quoted as relative to the measured
branching fraction. The uncertainty on the tag yield is estimated
from varying the background functions. Systematic uncertainties on
track finding and hadron particle identification efficiencies are
reported in Ref.~\cite{dhad}, while electron identification
efficiency is reported in Ref.~\cite{cleo-c-incl}. The sensitivity
to the requirement on the $e^+ \pi^-$ opening angle has been
evaluated by repeating the analysis after changing the requirement
by $\pm 5^\circ$. The model dependence of the efficiency is studied
using an alternative invariant mass distribution for the hadronic
system governed by phase space.  In the analysis where we apply a
mass requirement on the $K^-\pi^+\pi^-$ system, the model dependence
of the efficiency is estimated by varying the form factors in the
ISGW2 model, and the corresponding uncertainty is found to be 4\%.
The background uncertainty is derived by changing the measured fake
probabilities within their errors.

In summary, we have presented the first measurement of the absolute
branching fraction ${\cal B}(D^0\to K^-\pi^+\pi^-e^+\nu _e)=
[2.8^{+1.4}_{-1.1}\ \text{(stat)}\pm 0.3\ \text{(syst)}]\times
10^{-4}$. The invariant mass of the hadronic system recoiling
against the $e^+\nu_e$ pair is consistent with $\konem$. By
requiring $M_{\rm had}$ to be within the $[1150 - 1500]$ MeV/$c^{2}$
mass window, we obtain the product branching fraction ${\cal
B}(D^0\to \konem e^+\nu _e)\cdot({\cal B}(\konem\to K^-\pi^+\pi^-)=
[2.5^{+1.3}_{-1.0}\ \text{stat}\pm 0.2 ]\times 10^{-4}$. The
statistical significance is about 4.0 standard deviations. Using the
$\konem$ decay modes reported in the PDG \cite{pdg06}, we calculate
the $\konem \to K^-\pi^+\pi^-$ branching fraction to be (33 $\pm$
3)\%. Thus, the absolute branching fraction ${\cal B}(D^0\to\konem
e^+\nu _e)$ is $[7.6^{+4.1}_{-3.0}\ {\rm (stat)}\pm 0.6\ {\rm
(syst)} \pm 0.7]\times 10^{-4}$. The last error accounts for the
uncertainties in the measured $\konem$ branching fractions. This
channel is found to be 1.2\% of the total semileptonic width. The
ISGW \cite{isgw} model predicts this fraction to be about 1\%, while
the ISGW2 model \cite{isgw2} predicts this fraction to be about 2\%;
hence the measured branching fraction and $K^-\pi^+\pi^-$ invariant
mass are consistent with quark model calculations.


We gratefully acknowledge the effort of the CESR staff in providing
us with excellent luminosity and running conditions.
D.~Cronin-Hennessy and A.~Ryd thank the A.P.~Sloan Foundation. This
work was supported by the National Science Foundation, the U.S.
Department of Energy, and the Natural Sciences and Engineering
Research Council of Canada.

\end{document}